\documentstyle[aps,twocolumn]{revtex}

\newcommand{\re}{\mathop{\mathrm{Re}}}
\newcommand{\im}{\mathop{\mathrm{Im}}}
\newcommand{\tr}{\mathop{\mathrm{tr}}}
 
\newcommand{\dist}{\mathop{\mathrm{dist}}}
\newcommand{\diag}{\mathop{\mathrm{diag}}}

\title{Distribution of Eigenvalues in Non-Hermitian 
Anderson Model }
\author{Ilya Ya. Goldsheid and Boris A. Khoruzhenko}
\address{School of Mathematical Sciences, Queen Mary \& Westfield
College, \\ University of London, London E1 4NS, U.K.}
\date{22 July 1997}
\begin{document}
\draft
\maketitle
\bigskip

\begin{abstract}
We develop a theory which describes the behaviour of eigenvalues
of a class of one-dimensional random non-Hermitian operators 
introduced recently by Hatano and Nelson. Under general assumptions 
on random parameters we prove that the eigenvalues are distributed 
along a curve in the complex plane. An equation for the curve is 
derived and the density of complex eigenvalues is found in terms of 
spectral characteristics of a ``reference'' hermitian disordered 
system. Coexistence of the real and complex parts in the spectrum 
and other generic properties of the eigenvalue distribution for 
the non-Hermitian problem are discussed.  
\end{abstract}
\vfill

{\bf 1}. Complex eigenvalues of non-Hermitian random Hamiltonians 
have recently attracted much interest across several areas of 
physics \cite{HN,MWC,QCD,Yan,E}. However, the actual progress 
in understanding the statistics of such eigenvalues 
is mostly limited to the so-called `zero-dimensional' case,
i.e. to random matrices with no spatial structure. In this case
a number of models can be treated analytically and their basic 
features are relatively well understood, although on different 
levels of rigour (see, for instance, recent works \cite{Yan,E,RM} 
and references therein).  In contrast, little is known
about spectra of non-Hermitian Hamiltonians in one or more 
dimensions. One of the challenging problems here involves an 
unusual localisation-delocalisation transition predicted by 
Hatano and Nelson\cite{HN}.

Motivated by the studies of statistical mechanics of 
the magnetic flux lines in superconductors with columnar defects, 
Hatano and Nelson considered a model described by a random 
Schr\"odinger operator with a constant imaginary vector potential. 
Appealing to a qualitative reasoning they argued that already in 
one dimension some localised states undergo a delocalisation 
transition  when the magnitude of potential increases. 
The eigenvalues corresponding to the localised states are real 
and those corresponding to the extended states are complex. 
The results of numerical calculations presented in \cite{HN} 
support these conclusions. They also show a surprising feature of 
the  eigenvalue distribution in the model: 
the eigenvalues are attracted to a curve in the complex plane.   

In this short communication we explain this feature. We also derive 
an equation for the curve and obtain the density   
of complex eigenvalues in terms of the well known objects of 
the conventional (Hermitian) disordered systems.

Most of our discussion will involve the lattice case which 
is technically simpler. Following \cite{HN}, 
we will consider a one-dimensional non-Hermitian Anderson model 
whose eigenvalue equation reads as follows
\begin{eqnarray}\label{4}
-e^{\xi_{k-1}}\psi_{k-1} -e^{\eta_{k}} 
\psi_{k+1} +q_{k}\psi_{k} = z\psi_{k}, & & \;\; 1\le k \le n \\
\psi_{0}=\psi_{n}, \;\; \psi_{1}=\psi_{n+1}.\label{5}  
\end{eqnarray}
The relevance of this non-Hermitian Anderson model 
to physics of vortex lines in superconductors
is explained in\cite{HN}. In these works two particular cases 
of the model are discussed: (i) `site-random' model with random $q_k$ 
and with constant hopping elements  
given by $\eta_k \equiv h$ and $\xi_k \equiv -h$
($h$ is real); and 
(ii) `random-hopping' model  with $q_k\equiv 0$ and 
with random hopping
elements given by $\eta_k =g^{+}a_k $ and $\xi_k =g^{-}a_k$,
where $g^{\pm }$ are some constants and 
the $a_k$ are random variables. 

Our basic assumptions about the 
coefficients in Eq.\ (\ref{4}) are as follows: 
$\{ (q_k, \xi_k, \eta_k) \} $ is a stationary sequence of 
random three-dimensional vectors such that  
$\langle \log (1+|q_0|)\rangle $, 
$\langle \xi_0\rangle $, 
$\langle \eta_0\rangle $ are finite.
The angle brackets denote averaging over the disorder.  

\smallskip 

{\bf 2}. We start our analysis with a 
standard transformation which is often used in the 
theory of differential and difference equations. 
Let us put $\psi_k = w_k \varphi_k$ in 
Eq.\ (\ref{4}) and choose the weight $w_k$ 
so that to make the resulting equation symmetric.
For instance, if we set 
\begin{eqnarray}\label{d}
w_0=1 \; \mbox{and}\: 
w_k=e^{\frac{1}{2} \sum_{j=0}^{k-1} (\xi_j-\eta_j)}\;
\mbox{if}\; k \ge 1,   
\end{eqnarray}
then this transformation reduces the eigenvalue problem 
(\ref{4})-(\ref{5}) to the following one
(we will use the notation $c_k=\exp [(\xi_k+\eta_k)/2  ]$ onwards)
\begin{eqnarray}
\label{6}
-c_{k-1}\varphi_{k-1} -c_{k} \varphi_{k+1} 
+q_{k}\varphi_{k}=z\varphi_{k} \\
\label{bc}
\varphi_{n+1}= w_{n+1} ^{-1}w_1\varphi_{1},\;\; 
\varphi_{n}=w_n ^{-1}\varphi_{0}. 
\end{eqnarray}
From now on, we will deal with  Eqs.\ (\ref{6})--(\ref{bc}).
[Obviously, the eigenvalues of (\ref{4})--(\ref{5}) and 
(\ref{6})--(\ref{bc}) coincide.] 

We can now introduce a ``reference'' symmetric 
eigenvalue problem  which will be used in 
our analysis. This reference problem is specified 
by the same Eq.\ (\ref{6}) and with  the following 
boundary conditions (b.c.) 
\begin{eqnarray}\label{7}
\varphi_{n+1} = 0, \;\;   \varphi_{0} = 0.
\end{eqnarray}
One can  rewrite  the eigenvalue problems 
(\ref{6}),(\ref{bc}) and 
(\ref{6}),(\ref{7}) in the matrix form  
$ {\cal H}\psi = z \psi $ and 
$H \varphi = z \varphi $, respectively. 
We use the calligraphic ${\cal H}$ 
for the non-symmetric problem (\ref{6}),(\ref{bc}) 
and the usual 
italic $H$ for the symmetric
problem (\ref{6}),(\ref{7}).
$H$ is a symmetric tridiagonal $n\times n$ 
matrix with the $\{q_k\}$ on the  
main diagonal and the $\{-c_k\}$ on the sub-diagonals. 
${\cal H}$ is ``almost'' tridiagonal:
the only non-zero elements of the difference $V={\cal H} - H$ 
are $V_{1n} = -w_1^{-1}w_n e^{\xi_0} $ and 
$V_{n1} = -w_1w_n^{-1} e^{\eta_n}$.

Let $z_1, \ldots , z_n$  
denote the eigenvalues of 
${\cal H}$. Distribution of $\{z_j \}$ in the 
complex plane $z=x+iy$ is described by the measure  
\[ 
d\nu_n (z) \equiv d\nu_n (x,y) =\frac{1}{n} \sum_{j=1}^n
\delta (x-x_j) \delta (y-y_j) dx dy.
\] 
Our goal now is to find the limit of $d\nu_n (z)$
when $n\to \infty $. To do this we will calculate 
the ``electrostatic'' potential $F(z)\equiv F(x,y)$
of the limit distribution of the eigenvalues of ${\cal H}$:
\begin{eqnarray}\label{pot}
F (z) = 
\lim_{n\to \infty} F_n (z), 
\end{eqnarray}
where
\begin{eqnarray}\label{10a}
F_n(z)= \int_{\bf C}  
 \log |z-z'|  d\nu_n (z')=  \frac{1}{n}
 \log |\det ({\cal H}-zI )|
\end{eqnarray}
is the potential of $d\nu_n (z) $. In Eq.\ (\ref{10a}) $I$
is the $n\times n$ identity matrix. 

According to the potential theory, the existence 
of the limit in Eq.\ (\ref{pot})
for almost all $z$ implies the existence of the weak limit 
$ d\nu (z) = \lim_{n \to \infty }d\nu_n (z)$. The 
limit measure 
is determined by the Poisson equation, i.e.  
$ d\nu (z) \equiv 
d\nu (x,y) = 
1/(2\pi) \Delta F(x,y)\ dxdy$\cite{H,W}. 

The idea of using potentials to study eigenvalue
 distributions    
is not new and goes back to the 1960s at least, to studies 
of the eigenvalue distribution of T\"oplitz matrices \cite{W}. 
In the context of random matrices this idea
has been used since works \cite{Gir,Som}.

We will prove below that with probability one the limit 
in Eq.\ (\ref{pot}) exists for almost all $z$ and is given by  
\begin{eqnarray}\label{13}
F(z)=
\left\{
\begin{array}{l l}
a
& \mbox{if}\ \Phi (z) < a \\
\Phi (z) & \mbox{if}\ \Phi (z) > a, 
\end{array}
\right.
\end{eqnarray}
where $a= \max (\langle \xi_0 \rangle , 
\langle \eta_0 \rangle )$
and  $\Phi (z)$ is the potential of the limit
 distribution
of the eigenvalues of $H$. In other words,
$\Phi (z)=\lim_{n\to \infty} \Phi_n (z)$, where
\[
\Phi_n(z)=\frac{1}{n} \log |\det (H-zI)| =
\int_{-\infty}^{+\infty }\!\! \log |z-\lambda | dN_n(\lambda ) 
\]
and 
$N_n(\lambda )$ is the number of eigenvalues of $H$ 
in the interval $(-\infty , \lambda )$ divided by $n$. 
It is well known \cite{BL,CL} that there exists a continuous 
non-random 
function $N (\lambda )$ such that with probability one 
$\lim_{n\to \infty} N_n(\lambda )=N (\lambda )$ and hence
\begin{eqnarray}\label{12a}
\Phi (z) = 
\int_{-\infty}^{+\infty } \log |\lambda - z| dN(\lambda ).
\end{eqnarray} 

We will call the set which supports 
$d\nu (z)$ the limit spectrum of ${\cal H}$.    
$\Phi (z)$ as a function of $x=\re z$ and $y=\im z$ 
is harmonic everywhere in the complex 
plane except that part of the real 
axis which supports 
$dN(\lambda )$. Therefore, Eq.\ (\ref{13})  says that
the complex part of the limit spectrum 
is determined by the equation 
$\Phi (z) =a $. This equation defines a curve 
in the complex plane. We denote this curve by ${\cal L}$. 
The density of the eigenvalue distribution 
with respect to the 
arc-length measure $ds$ on this curve 
equals to $(2\pi)^{-1}$ times jump in the normal derivative of
$F$ across the curve. Computing the derivative gives 
\begin{eqnarray}\label{13a}
\frac{d\nu }{ds}=\frac{1}{2\pi}\left| \int_{-\infty }^{+\infty} 
\frac{dN(\lambda )}{\lambda -z}
 \right|,\;\;  z\in {\cal L}. 
\end{eqnarray}

$N (\lambda )$ is the classical integrated density of states 
associated with Eq. (\ref{6}).
It is very important that $\Phi (z)$ 
coincides,  up to an additive constant, with   
the Lyapunov exponent $\gamma (z)$ of the same equation. Namely,
the well known Thouless formula\cite{HJT} 
states that
$
\Phi (z)=\gamma (z) + \frac{1}{2}
(\langle \xi_0 \rangle + \langle \eta_0 \rangle )
$
(for a rigorous proof of this formula and  
further references see e.g. \cite{PF}).

The spectral properties of ${\cal H}$ 
are, to a large extent, determined 
by the behaviour of $\gamma (z)$. This will be discussed later.
Here we only mention that the equation $\Phi (z)=a$ which defines
the curve ${\cal L}$ is equivalent to 
\begin{eqnarray}\label{13b}
\gamma (z)=
\frac{1}{2}|\langle \xi_0 \rangle - \langle \eta_0 \rangle  |.
\end{eqnarray}
Eq.\ (\ref{13b}) is more transparent. Loosely speaking, 
it says that a non-real $z$ belongs to the 
limit spectrum  of ${\cal H}$ only if there exists 
a solution $\varphi $
 of  Eq.\ (\ref{6}) such that $|\varphi_n |$ and 
$w_n^{-1}$ are of the 
same order when $n \to \infty $ [compare this with
 Eq.\ (\ref{bc})]. 
As an immediate consequence of Eq. (\ref{13b}) we 
obtain the following.
If $\langle \xi_0 \rangle = \langle \eta_0 \rangle $ 
then 
there is no complex part in the limit spectrum of 
${\cal H}$, i.e.
$d\nu (z)$ is supported on the 
real axis only. 
 
To derive 
Eq.\ (\ref{13}) let us consider a non-real $z$
such that $\Phi (z) \not= a$. Then $G=(H-zI)^{-1}$ 
is well defined and ${\cal H} -zI= (I+VG)(H-zI)$. 
Therefore
\begin{eqnarray*}
F_n (z)   
 = \Phi_n(z)+\frac{1}{n} \log |d_n(z)|, 
\end{eqnarray*}
where $d_n(z)= \det (I+VG) $. Expanding this determinant 
yields
\[
d_n(z)=[1+V_{1n} G_{n1}][1+V_{n1} G_{1n}]- V_{1n}V_{n1}
G_{11}G_{nn},  
\]
where $G_{jk}$ denotes the $(j,k)$ matrix entry of $G$. 
To prove Eq.\ (\ref{13}) it remains to find  the
$ \lim_{n\to \infty} n^{-1}\log |d_n(z)|$.

Using $G_{1n}=G_{n1}=\prod_{j=1}^{n-1} c_k /\det(H_0-zI)$ 
one obtains
\begin{eqnarray*}
|V_{1n} G_{n1}|&=&\exp \{ n[\langle \xi_0 \rangle 
-\Phi (z) +r_n(z)+s_n]\},\\
|V_{n1} G_{1n}|&=&\exp \{ n[\langle \eta_0 \rangle 
-\Phi (z) +r_n(z)+t_n]\},
\end{eqnarray*}
where $r_n(z)=\Phi (z)-\Phi_n (z)$, $s_n=n^{-1}
\sum_{j=0}^{n-1}\xi_j -
\langle \xi_0 \rangle$, and 
$t_n=n^{-1}\sum_{j=1}^{n} \eta_j - \langle \eta_0 \rangle $. 
As mentioned above, $\Phi_n (z) \to \Phi (z) $ 
and hence $r_n (z) \to 0$  when $n\to \infty $. Also, because of 
the ergodic
theorem $\lim_{n \to \infty }s_n = \lim_{n \to \infty }t_n =0$.

Consider first the case when $\Phi(z) < a$. 
Without loss of generality we may suppose that 
$a  
= \langle \eta_0 \rangle $. Then $V_{n1}G_{1n}$ grows  
exponentially with $n$. 
On the other hand, 
$\Phi(z) > \langle \xi_0 \rangle $ for every non-real $z$\cite{fn1}  
and $V_{1n}G_{n1}$ tends to 
zero exponentially fast when $n\to \infty $. Besides, 
$V_{1n}V_{n1}=\exp [(\xi_0 + \eta_n )/2]$ and the stationarity 
of the sequence $\{\eta_n \}$  implies that 
\begin{eqnarray}  \label{12e}
\lim_{n\to\infty}n^{-1}\log |V_{1n}V_{n1}| = 0.
\end{eqnarray}
Therefore 
$\lim_{n\to\infty}n^{-1}\log |d_n (z)|= a-\Phi(z)$,  
hence $\lim_{n\to \infty } F_n(z) = a$ if $\Phi (z) < a$.

Consider now the case when  $\Phi(z) > a$. Then both 
$V_{n1}G_{1n}$ and $V_{1n}G_{n1}$ are exponentially small
if $n$ is large. Therefore the ``dangerous'' term in $d_n (z)$ 
is $1-V_{1n}V_{n1}G_{11}G_{nn}$.  
It can be shown\cite{GK} that $\alpha \le \arg G_{11}
 \le \pi - \alpha$, 
where $\alpha $ is strictly positive ($0<\alpha < \pi/2)$ 
and depends only on $z$ 
and on the realisation of $\{ (q_k, \xi_k, \eta_k ) \}$.
One can deduce from this inequality that
\[
|1-V_{1n}V_{n1}G_{11}G_{nn}|\ge \sin \alpha .
\] 
This estimate together with Eq.\ (\ref{12e}) and 
the simple inequality 
$|G_{jj}|\le |\im z|^{-1}$ prove
that $n^{-1}\log |d_n (z)| \to 0$ when $n \to \infty $.
Thus 
$\lim_{n\to \infty } F_n(z) = \Phi (z)$ if $\Phi (z) > a$.

Since the limit distribution of eigenvalues of ${\cal H}$ is 
one-dimensional, one can also use the conventional technique
of Green's functions to obtain this limit distribution.
Let $z$ be non-real and $f_n (z)=n^{-1}\tr ({\cal H}-zI)^{-1}$, 
$g_n (z)=n^{-1}\tr (H-zI)^{-1}$, 
$f(z)=\lim_{n\to \infty }  f_n (z)$,
and $g(z)=\lim_{n\to \infty } g_n (z)$. 
Revising the above argument 
one can show that $f(z)=0$ if $\Phi (z) < a$ and
$f(z)=g(z)$ if $\Phi (z) > a$. This can also be shown by working 
directly with the Green's functions $f_n (z)$ and $g_n (z)$.

If $\{(q_k, \xi_k, \eta_k )\}$ is a sequence of 
independent identically distributed 
random vectors with finite variance   
($\xi_k$ and $\eta_k$ are not
necessarily independent), then  
more precise information 
on the large-$n$ behaviour of $r_n(z)$, $s_n$, and $t_n$  
is available.  
Revising the above analysis and employing the limit-theorem 
techniques 
one can show in this case that 
\begin{eqnarray}\label{dist}
\rho_n \equiv\max\limits_{z_j: 
|\im z_j | > \varepsilon } \dist (z_j, {\cal L})
\end{eqnarray}
is of the order $1/\sqrt{n}$ for any positive $\varepsilon $. 

\smallskip 

{\bf 3}. In this section, we consider another approach to the 
eigenvalue 
problem (\ref{6})--(\ref{bc}). It  is based on the theory of  
Lyapunov exponents of 
products of random matrices. 
This approach explains the true reason for the appearance 
of the complex eigenvalues and gives 
an independent derivation of Eq. (\ref{13b}). It also 
provides more precise 
information about the finite-$n$ behaviour of the eigenvalues 
of ${\cal H}$.  Finally, it easily extends to the case of 
differential equations. It does not provide
though the explicit form of the limit distribution 
derived above.

Let us introduce the usual transfer-matrices
\begin{eqnarray}\label{tm}
A_k =\frac{1}{c_k}
\left( 
\begin{array}{c c }
q_{k}-z  & -c_{k-1}    \\
c_k      &  0    \\
\end{array}
\right).
\end{eqnarray}
Then the solution of Eq.\ (\ref{6}) with 
initial data $(\varphi_0 , \varphi_1)$
can be written as 
\begin{eqnarray}\label{pm}
(\varphi_{k+1}, \varphi_{k})^T=S_k(z)(\varphi_{1}, 
\varphi_{0})^T,
\end{eqnarray}
where $S_k(z)= A_{k}A_{k-1}\ldots A_{1} $.  
Thus the eigenvalue problem  (\ref{6}) -- (\ref{bc})
reduces to 
\begin{eqnarray}\label{em}
\left(B_n S_n(z)-w_{n+1}^{-1}I \right)(\varphi_{1}, 
\varphi_{0})^{T}
 =0,
\end{eqnarray}
where $I$ is the $2\times 2$ identity matrix and $B_n$ is a 
 $2\times 2$ diagonal matrix, 
$B_n=\diag \{\exp [-\frac{1}{2}(\xi_0 - \eta_0)], 
\exp [-\frac{1}{2}(\xi_n - \eta_n)] \}$.
In other words, the eigenvalues of ${\cal H}$ solve the 
equation $\det (B_n S_n(z)-w_{n+1}^{-1}I) =0$.
This equation is equivalent to $d_n(z)=0$. 

The Lyapunov exponent $\gamma (z)$ of  
$S_n(z)$ is defined as 
\begin{eqnarray}\label{le}
\gamma(z)=\lim_{n \to \infty} \frac{1}{n}\log || S_n (z)||.
\end{eqnarray}
It is well known 
(\cite{F}, \cite{BL} ) that under our basic assumptions 
on $\{ (q_k, \xi_k, \eta_k )\}$ the limit on the right-hand side
in Eq.\ (\ref{le}) exists with probability one.

For every fixed $z$ the right-hand side of Eq.\ (\ref{le})
gives with probability 1 the same non-random number
as the Thouless formula and the well known Furstenberg's formula\cite{F}.
To proceed, however, we have to study $ \gamma (z)$ as a
function of $z$ when a (typical) realisation of  
$\omega\equiv \{ (q_k, \xi_k,\eta_k )\}$ is fixed:
$ \gamma (z)= \gamma (z, \omega)$.
Thus, we are faced with products of random matrices depending 
on a parameter. The latter were studied in \cite{G}. 

Let $\Sigma$ denote the spectrum 
the operator defined by the left hand side of Eq.\ (\ref{6}) on 
$l_2(\bf Z)$ [this set can equivalently be described as the 
support of the measure $dN(\lambda)$]. 
It was proved in \cite{G} 
that with probability 1
the convergence in (\ref{le}) is uniform in $z$ if $|z|\le C$ and
$ z $ does not belong to an $\varepsilon$-neighbourhood of 
$\Sigma$;
here $C$ is an arbitrary fixed large number and $\varepsilon$ is 
an arbitrary 
strictly positive number.
It can be deduced from this fact that:
(a) if $\mu_n(z)$ is the largest by modulus eigenvalue 
of $B_nS_n$  then $ \lim_{n \to \infty}  n^{-1}\log|\mu_n(z)| = 
\gamma(z)$ and 
the 
convergence here is again uniform in $z$ in the same 
domain (clearly, 
the other eigenvalue is 
$e^{\xi_n+\eta_0}/\mu_n(z)$).
(b) The following function $\bar\gamma(x) \equiv 
\gamma(x+i0) $
is now defined for almost every sequence $\omega$ 
simultaneously 
for all real $x$, $-\infty<x<\infty$;
the latter limit exists because $\gamma(z)$ is monotone in 
$y$ (the property which is obvious from the Thouless formula).

There is a subtlety here. Namely, we can consider $\gamma(x)$ 
defined by (\ref{le}) for every fixed(!)  $x$ but not 
for all $x$  simultaneously while the potential is fixed. Moreover, 
it turns out
that if a typical sequence $\omega $ is fixed then 
there exists a "large" subset $\Theta_\omega$ of  
$\Sigma $ such that
for all $x\in \Theta_\omega$ the limit in (\ref{le})
 either does not exist 
 or 
 does not coincide with $\bar\gamma(x)$ \cite{G}. 
However the following 
statement which is very important for us holds true:
\begin{eqnarray}\label{us}
\bar\gamma(x)\ge \lim_{n \to \infty}\sup
 \frac{1}{n}\log || S_n(x) ||\equiv\tilde\gamma(x)
\end{eqnarray}
A more detailed discussion of properties (a), (b), 
and (\ref{us})
along with a very simple proof of the mentioned 
above uniform convergence 
in (\ref{le}) will be provided in \cite{GK}.

Combining property (a) with the statement that
$\lim_{n \to \infty} n^{-1} \log w_{n+1}= 
(\langle \xi_0 \rangle - \langle \eta_0 \rangle )/2 $
one concludes that all non-real solutions of Eq.\ 
(\ref{em}) are asymptotically 
(as $n\to\infty$) attracted to the curve $\cal L$ 
given by Eq.\ (\ref{13b}).
Moreover, it follows from the uniform convergence 
of  $\mu_n(z)$ 
that $ \lim_{n \to \infty}\rho_n = 0$, where $\rho_n$ 
is defined by (\ref{dist}).

It is worth noticing that Eq.\  (\ref{le}) can very 
rarely be used for calculation
of the Lyapunov exponent $\gamma (z)$. 
In order to make use of Eq.\ (\ref{13b})
one has to compute $\gamma (z)$. One way to do this is provided 
by  the Thouless formula. The other one is the well
known Furstenberg's formula\cite{F}. To a large extent, 
both approaches 
are equivalent and in order to use any of these formulae
one has to find an invariant measure
of a certain transformation for which purpose
an integral equation has to be solved. Though in general 
calculations of 
this sort are
difficult, for some distributions the explicit expression for 
$\gamma (z)$ and $N(\lambda )$ can be found; one of these examples
is discussed below. 

The properties of the two functions, $\bar\gamma(x)$ 
and $\tilde\gamma(x)$,
are crucial for further analysis. 
Suppose for simplicity that in addition to our basic 
assumptions
$ (q_n,\xi_n,\eta_n)$ form a sequence of independent 
identically 
distributed random vectors. Then it is easy to show that
for sufficiently large $|x|$ 
\begin{eqnarray}\label{bound}
\log |x| -C_1 \le \bar\gamma(x)\le \log |x |+C_1,
\end{eqnarray} 
where $C_1$ depends only on the distribution of 
$ (q_0,\xi_0,\eta_0)$.
Then, obviously, there exist $C$ such that all solutions of 
Eq.\  (\ref{13b}) belong to a circle $|z|\le C$.

The other useful property of $\bar\gamma(x)$ is as follows: 
if in addition to the
above assumptions $ q_0 $ takes at least two different 
values then 
$\bar\gamma(x)>0$ is strictly positive (see \cite{F,CL,PF}). 
Finally, under these
conditions, $\bar\gamma(x)$ is a continuous function of 
$x$ (see, e.g. \cite{G}).

We are now in a position to describe the curve ${\cal L}$. 
Consider the inequality 
\begin{eqnarray}\label{zz}
\bar\gamma(x) \le \frac{1}{2}
|\langle \xi_0 \rangle - \langle \eta_0 
\rangle  |   
\end{eqnarray}
Because of the continuity of $\bar\gamma(x)$, 
the solution of this inequality 
is given by a union of  disjoint intervals:  
$\cup_j[a_j,a_j^\prime]$ with $ a_j<a_j^\prime$.
Next,  for every $x \in [a_j,a_j^\prime ]$, 
consider a positive $y =y_j(x)$ such that 
the pair $(x,y)$ solves Eq.\ (\ref{13b}). This 
solution exists and is unique 
because if $x$ is fixed then $\gamma(x+iy)$ is 
strictly monotone continuous
function of positive $y$ and $\lim_{y\to +\infty}\gamma(x+iy)
=\infty$.
The curve ${\cal L}$  (the solution of Eq.\  
(\ref{13b})) 
is a union of disconnected contours: ${\cal L}=\cup_j {\cal L}_ j$. 
Each contour ${ \cal L}_j$ consists of two symmetric smooth arcs
$y=y_j(x)$ and $y=-y_j(x)$, where $a_j \le x \le a_j'$;
the points $(a_j, 0)$ and 
$( a_j^\prime,0)$  are the end-points of these  arcs.
[We notice that  it is easy to construct examples 
with a prescribed finite number of contours. 
In general, there is no obvious reason for the 
number of contours to be finite for 
an arbitrary distribution of $(\xi_0,\ \eta_0,\ q_0)$.]

Our next goal is to describe the behaviour of the real
eigenvalues of Eqs.\ (\ref{6})--(\ref{bc}).
It turns out that this behaviour is governed by the 
following remarkable
property of  $\tilde\gamma(x)$: 
this function is upper semi-continuous at each point 
$x$ where
$\tilde\gamma(x)= \bar\gamma(x)$ (this equality holds 
for almost every $x$;
remember that $\tilde\gamma(x) $ depends strongly on 
$\omega$ \cite{G})
 
It can be deduced from this property \cite{GK} that 
for every strictly positive 
$\varepsilon$ the spectrum of ${\cal H}$  
lies outside a domain
surrounded by the following strip:
\[
{\cal D}_{j, \varepsilon }\equiv \{z\in {\bf C}:
\dist( |z|,{\cal L}_ j) \le\varepsilon \}
\]
if $n$ is large enough.
In other words, with probability 1 the spectrum of  
${\cal H}$  is wiped 
out from
the interior of every contour ${\cal L}_ j$ 
as $n\to\infty$.

Finally, here is a description of the limit  spectrum of Eqs.\
(\ref{6})--(\ref{bc}).
Let $\Sigma$ be the same as above (i.e. the spectrum of 
the reference symmetric problem). Then the set 
\begin{eqnarray}\label{sp}
{\cal L} \cup \left\{ \Sigma \backslash \cup_j(a_j,a_j^\prime )\right\}
\end{eqnarray}
is the limit spectrum of ${\cal H}$.

 \smallskip 

{\bf 4}. We start our discussion with a remark on the 
spectrum of the limit operator $\hat {\cal H}$  defined 
by the left-hand side of Eq.\  (\ref{4}) on $l_2 ({\bf Z})$.
For simplicity we suppose that all coefficients in Eq.\ 
(\ref{4}) are bounded. It turns out that for a wide class
of distributions of $\{(\xi_k, \eta_k, q_k ) \}$ the spectrum of 
 $\hat {\cal H}$ is a two-dimensional subset in the
complex plane and the limit spectrum of ${\cal H}$ 
given by Eq. (\ref{sp}) is embedded into this set.  
This phenomenon seems to be surprising  because ${\cal H}\varphi $
converges to $\hat {\cal H}\varphi $ 
when $n \to \infty $ ($n$ is the dimension of ${\cal H}$) 
for every $\varphi \in l_2 ({\bf Z})$. 

Since the problem is non-Hermitian, the spectrum may depend on 
the choice of boundary conditions (b.c.). Indeed, all the 
eigenvalues of Eq. (\ref{4}) with the Dirichlet b.c., i.e. 
when $\psi_{n+1}=\psi_0=0$, are real. It is remarkable, however,  
that the boundary conditions of the form $(\psi_{n+1}, 
\psi_n)^T=B (\psi_{1}, \psi_0)^T $, where $B$ is a fixed 
real {\em non-degenerate} $2\times2$ matrix, lead to the 
same Eq.\ (\ref{13b}) (regardless of the choice of $B$) 
in the limit $n \to \infty $. For diagonal $B$ this fact can be 
readily seen from our proof of 
Eq.\ (\ref{13}).  The general case of a non-degenerate 
$B$ is less transparent and our proof of Eq.\ (\ref{13b})
in this case \cite{GK} relies on the 
above mentioned properties of products of random 
matrices\cite{G}. 

The reader might have noticed that our derivation
of Eq.\ (\ref{13}) is based only on the fact of existence 
of $N(\lambda )$. (Under our basic assumptions
the existence of $N(\lambda )$ is ensured by the stationarity of the 
coefficients.) Thus one can easily extend our argument 
to other classes of coefficients.
  
For instance, instead of a random sequence $\{(\xi_k,\eta_k,q_k)\}$ 
one can take a periodic one (i.e. the $q_k$, $\xi_k$, 
and $\eta_k$ all have a common period). Then Eq.\ (\ref{13}) 
holds with the following obvious change. In the periodic case 
the angle brackets denote averaging over the period.
The complex part of the limit spectrum is described by the 
same equation (\ref{13b}) in which $\gamma (z)$ 
is the Lyapunov exponent of the symmetric reference 
equation (\ref{6}) whose coefficients 
are now periodic.    

The geometry of the limit spectrum of ${\cal H}$ 
in the periodic case follows two patterns.
If $\langle \xi \rangle = \langle \eta \rangle$ then the 
limit spectrum is real and coincides with $\Sigma $, the 
spectrum of the symmetric reference equation. If 
$\langle \xi \rangle \not= \langle \eta \rangle$ the 
limit spectrum is purely complex, 
i.e. the limit distribution of eigenvalues 
is supported on the curve defined by 
Eq.\ (\ref{13b}). It is worth mentioning that in 
the periodic case this curve is a union of 
a finite number of analytic 
contours. Indeed, because  $\gamma (z)=0$ on $\Sigma $ 
in the periodic case, 
the above mentioned arcs join up  
smoothly.  In either case (real or complex spectrum) 
the corresponding
eigenfunctions are extended. 
The case when $\{(\xi_k,\eta_k,q_k)\}$ is quasi-periodic 
is much more delicate 
and deserves a separate study.

The rest of our discussion is based on known 
results on Hermitian 
disordered systems and the formulae derived above. 
From now on we assume that
the $\{q_k\}$ is a sequence of independent 
identically distributed random
 variables 
which is also independent of the $\{c_k\}$.
It is instructive to consider first an exactly solvable model. 

Let $c_k\equiv 1$ and $q_k$ are Cauchy distributed, i.e. 
Prob$\{q_k\in \Delta\}=\pi^{-1}\int_\Delta dq\ b / (q^2 +b^2)$.
In this special case an explicit expression for 
$N(\lambda )$ is known (see, e.g. \cite{PF}). 
Using this expression 
and the Thouless formula
one obtains that 
\begin{eqnarray}\label{41}
4\cosh \gamma (z)& = &
 \sqrt{(x + 2)^2+(b + |y|)^2} + \\ 
 & & 
\sqrt{(x - 2)^2+(b + |y|)^2} \nonumber , 
\end{eqnarray} 
where $x\equiv \re z$ and $y\equiv \im z$.
One can use this result to obtain the 
limit spectrum of a family
of non-Hermitian operators ${\cal H}$ (\ref{4})--(\ref{5}). 
Suppose that in Eq.\  (\ref{4}) $\eta_k =-\xi_k$ 
for every $k$ (e.g. one can take $\eta_k \equiv h$ and 
$\xi_k \equiv -h $ as in the site-random model).
Then in the reference equation $c_k\equiv 1$ and 
by straightforward 
computations involving Eqs.\ (\ref{13b}) and (\ref{41}) 
one finds that 
the complex part of the limit spectrum of 
${\cal H}$ consists of the two arcs 
\begin{eqnarray}
\nonumber
y(x)=\pm  
\left[ 
\sqrt{
\frac{(K^2-4)(K^2-x^2)}{K^2}
     } -b 
\right],
-x_b\le x \le x_b ,\\
 \label{zzz} 
\end{eqnarray} 
 where $K=2\cosh \ \langle \eta_0 \rangle $ 
and $x_b$ is determined by the condition $y(x_b)=0$. 
The real eigenvalues of ${\cal H}$ 
are distributed (in the limit $n \to \infty $) 
in the intervals $(-\infty , -x_b)$ and $(x_b, +\infty )$
with density equal to the density of eigenvalues of 
Eq.\ (\ref{6}) in these intervals.   

The $b$-dependence of the complex part of the limit spectrum 
is remarkably simple. 
If $b = 0$ the arcs (\ref{zzz}) form 
the ellipse $x^2/K^2+y^2/(K^2-4)=1$.
As $b $ increases each of the arcs moves 
(by translation in the $y$-direction) towards the real axis 
and reduces in length. At $b = \sqrt{K^2-4}$ the arcs 
(hence the complex part of the limit spectrum) disappear. 
In other words,
if $K < K_{cr}=\sqrt{4+b^2}$ Eq.\ (\ref{13b}) cannot be solved 
and the limit spectrum of ${\cal H}$ is real. 
If $K> K_{cr}$ then the limit spectrum of ${\cal H}$ has 
always two branching points from which the complex branches
grow.   

We should mention another example of the reference 
eigenvalue problem 
(\ref{6}) amenable to 
analytic treatment. In this example  
the $c_k$ are independent exponentially  distributed 
random variables and $q_k \equiv 0$. 
Thus this example is related to 
the random-hopping
non-Hermitian Anderson model. 
The known expression for $\gamma (z)$\cite{D}   
is not that simple as in the previous case 
and in order to analyse Eq.\ (\ref{13b}) 
one has to employ certain  approximations. 
This analysis goes beyond the scope of the 
present publication.

The next part of our discussion involves 
general analysis of Eqs.\ (\ref{13}) 
and (\ref{13b}). Denote 
$ 2g  \equiv\langle \eta_0 \rangle 
- \langle \xi_0 \rangle $. 
As mentioned above the limit spectrum of 
${\cal H}$ is entirely real if $g=0$. First, we want
to show that in the case when $q_k\equiv 0$
(random hopping model) the limit spectrum has a non-trivial complex part
for every non-zero $g$. Indeed, in this case
the transfer matrices $A_k$  (\ref{tm}) 
have  zero diagonal elements when $z=0$. 
Hence their products and the norm of  $S_n(0)$  can be 
easily computed (see, e.g. 
\cite{PF}). From this computation and the stationarity of 
the $c_k$ it  
follows easily that $\gamma (0)=0$. Therefore, if $g\not= 0$ 
the solutions to the equation $\gamma (z)=|g|$
constitute a non-trivial (non-real) curve. 
 
This is not the case if the distribution of $q_0$ (hence of all  $q_k$) 
is non-degenerate
(i.e. $q_0$ takes at least two different values). Then
there exists $g^{(1)}_{cr}>0$  
 such that for all $|g| \le g^{(1)}_{cr}$ the limit 
spectrum 
of ${\cal H}$ has no complex part. 
Indeed, if the $q_k$ are non-degenerate 
$\bar \gamma (x )\equiv \gamma (x+i0)$ is strictly 
 positive on 
the limit spectrum $\Sigma$ of the reference eigenvalue problem
(\ref{6}),(\ref{bc}). The  continuity  
of $\bar \gamma (z)$, Eq.\ (\ref{bound}) and the inequality 
$\gamma (x+iy) \ge \bar \gamma (x)$ imply that 
\[
g^{(1)}_{cr}\equiv  \min_{x\in \Sigma} \bar \gamma (x ) >0.
\]
If in addition the coefficients in Eq.\ (\ref{6}) are bounded,
i.e. $c_k^2+q_k^2 \le C$ for all $k$, then 
$\Sigma$ is a bounded set. Therefore
\[
g^{(2)}_{cr} \equiv \max_{x \in \Sigma} 
\bar \gamma (x)  
\]
is finite. If $|g|\ge  g^{(2)}_{cr}$ the inequality  
(\ref{zz}) is satisfied for every point of $\Sigma $.
Hence the limit  spectrum is purely complex in this case. 
If either $c_0$ or $q_0$ takes 
arbitrary large values with non-zero probability, then
$\Sigma$ is an unbounded set and in view of Eq.\ (\ref{bound})
$g^{(2)}_{cr}=+\infty $. 

In summary, if $|g| \le g^{(1)}_{cr}$
the limit spectrum of ${\cal H}$ is entirely real, 
if $|g| \ge  g^{(2)}_{cr}$ the limit spectrum is entirely complex 
and if $g^{(1)}_{cr} < |g| < g^{(2)}_{cr}$ the limit spectrum has 
real and complex parts.  In the latter case the branching points
from which the complex branches grow out of the real eigenvalues are 
determined by $\bar \gamma (x)=|g|$. 

The limit distribution of the real eigenvalues of 
${\cal H}$ is described by $N(\lambda )$, the integrated 
density of states of the reference symmetric equation. 
The limit distribution of the non-real eigenvalues of 
${\cal H}$ is described  by Eq.\ (\ref{13a}). 
It should be noticed that the density of the non-real 
eigenvalues given by Eq.\ (\ref{13a}) is analytic everywhere 
on ${\cal L}$ except the (real) end-points of the arcs. 
(If the limit spectrum is entirely complex then this density 
is analytic everywhere.)  The behaviour of the density of 
the non-real eigenvalues near an end-point $x_j$ of an arc  
depends on the regularity properties of $N(\lambda )$ 
at this point. If the density of states $N'(\lambda )$ of the 
reference equation is smooth in a neighbourhood of 
the point $\lambda =x_j$ 
then the density of the limit distribution of the 
non-real eigenvalues of ${\cal H}$  
has a finite limit as $z$ approaches $x_j$ along the arc. 
Also, in this case the tangent to the arc exists at $x_j$ 
and is not vertical. In other words, if $N'(\lambda )$ is smooth 
in a neighbourhood of a branching point $\lambda = x_j$   
the complex branches of the limit spectrum grow out  
of $x_j$ linearly. This may not be the case if 
$N'(\lambda )$ is not smooth.

After this work was completed we learned about recent  
preprints \cite{r} addressing similar problems.


\begin{references}
\bibitem{fn1} This inequality can be inferred from 
the Thouless formula and the fact that $\gamma(z)\ge 0$ 
for real $z$.
\end{references}
\end{document}